\documentclass[runningheads]{llncs}
\usepackage[T1]{fontenc}
\usepackage{tabularx}
\usepackage{graphicx}
\newcommand{\add}[1]{{#1}}

\begin{document}
\title{The Future of Fully Homomorphic Encryption System:
from a Storage I/O Perspective}
\titlerunning{The Future of FHE System}
%
\author{Lei Chen\inst{1} \and Erci Xu\inst{2} \and Yiming Sun\inst{2} \and Shengyu Fan\inst{3}
\and Xianglong Deng\inst{3} \and Guiming Shi\inst{4} \and Guang Fan\inst{1} \and Liang Kong\inst{1} \and Yilan Zhu\inst{1} \and Shoumeng Yan\inst{1} \and Mingzhe Zhang$^*$\inst{1}
}
\authorrunning{L. Chen et al.}
%
\institute{Ant Group, China \and
Shanghai Jiaotong University, China
 \and
University of Chinese Academy of Sciences, China 
\and 
Tsinghua University, China
}
\maketitle              
\begin{abstract}
Fully Homomorphic Encryption (FHE) allows computations to be performed on encrypted data, significantly enhancing user privacy. However, the I/O challenges associated with deploying FHE applications remains understudied. 
We analyze the impact of storage I/O on the performance of FHE applications and summarize key lessons from the status quo. Key results include that storage I/O can degrade the performance of ASICs by as much as 357$\times$ and reduce GPUs performance by up to 22$\times$.$\footnotetext[1]{Corresponding author. Email: huayi.zmz@antgroup.com}$

\end{abstract}
%
%
%

\section{Introduction}

Security is always pivotal, especially for users who run applications with sensitive data on third-party platforms such as the cloud. Fully Homomorphic Encryption (FHE) provides a promising solution by enabling operations to directly perform on encrypted data. The user can decrypt the results which would be the same as if operations were conducted on plain text (i.e., unencrypted data). This allows users to offload computation on confidential data to public clouds for better performance without security concerns.

\begin{figure}[htbp]
\centering
    \includegraphics[width=0.85\linewidth]{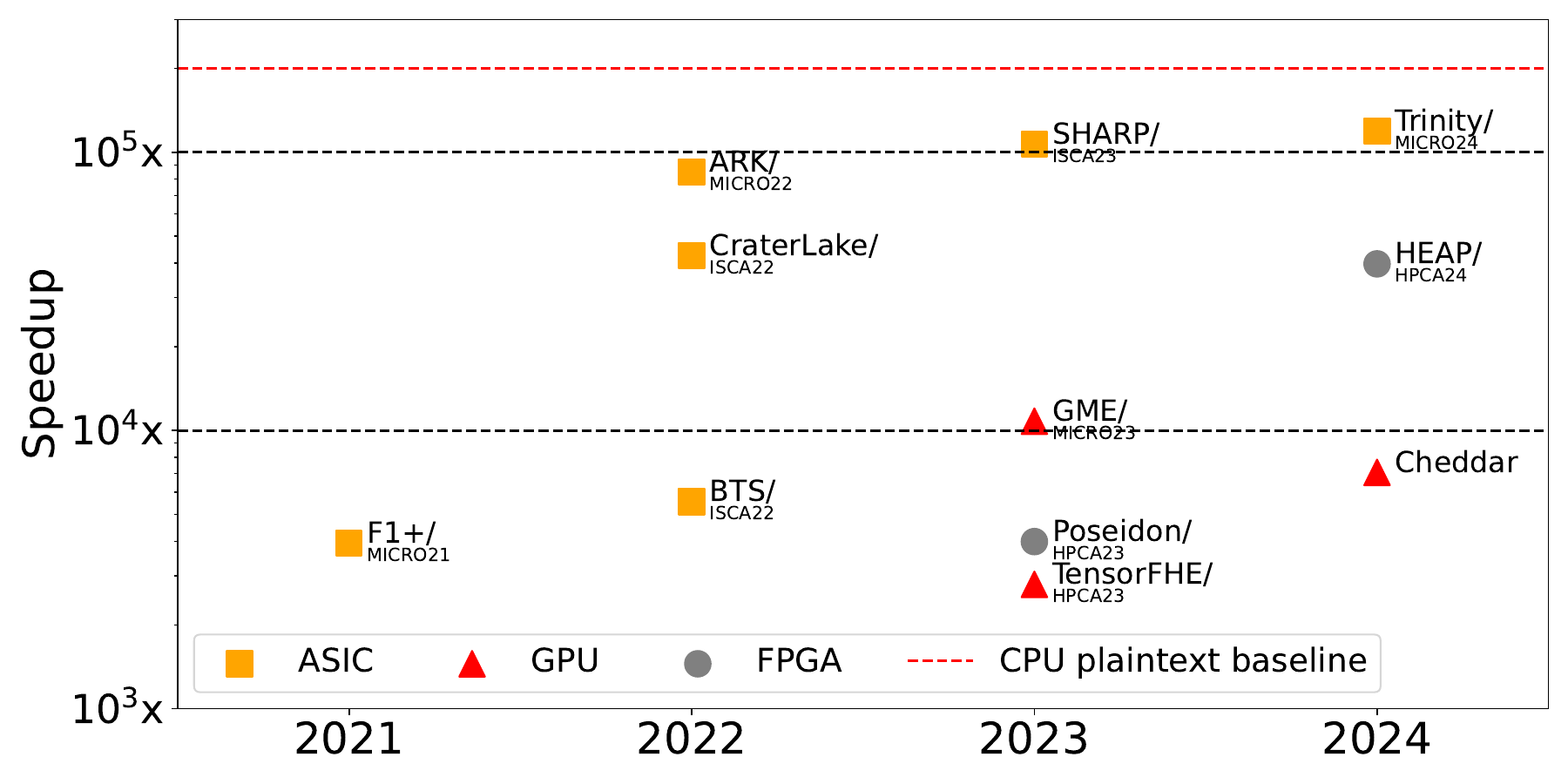} 
\caption{Speedup of FHE accelerators.\label{fig:fhe-paper}}
\end{figure}

However, the lunch is not free. Previous research reveals that achieving FHE can take a heavy toll on computation. Due to \add{cryptographic security requirements},  FHE computation on CPUs is more than $10^5\times$ slower than performing on plain text~\cite{ckks-bootstrapping}. Thus, a spate of works has been focusing on speeding up the computation by using co-processing units such as GPU~\cite{tensorfhe@hpca23,gme@micro23,cheddar} and ASIC~\cite{mad@micro23,bts@isca22,ark@micro22,sharp@isca23,craterlake@isca22,f1@micro21,poseidon@hpca23,trinity@micro24}. Fig.~\ref{fig:fhe-paper} shows the performance speedups of different FHE accelerators compared to computing FHE on the CPU. Latest prototypes~\cite{sharp@isca23,trinity@micro24} have shown that the gap is quickly diminishing with only a $3\times$ difference in processing. Moreover, more advanced solutions that can combine multiple FHE schemes to improve the performance further are looming on the horizon~\cite{heap@isca24}.

Now, the future of FHE seems bright except for one aspect, storage I/O, left overlooked. Prior works often are based on two assumptions of I/O accessing: (1) \add{all data is stored in the high-bandwidth device memory}; and (2) \add{the overhead associated with fetching data from the device memory to on-chip cache can be eliminated}. Unfortunately, these two may not hold, especially under the multi-user environment like the cloud. First, \add{each user has their own ciphertexts and evaluation keys (see Section~\ref{sec:background} ), and these data themselves may exceed the size of the device memory}. Second, \add{a server may provide services for multiple users, and it is impossible to keep all the data of all users in the device memory}.

To quantitatively evaluate the potential impacts of I/O on deploying FHE in the cloud, we conduct an extensive set of experiments using SimGrid~\cite{simgrid}. By testing representative FHE applications including Resnet-20~\cite{fhe-resent} and HELR~\cite{helr} with varying storage devices and networks, the results motivate us to gather three important lessons:
\begin{itemize}
    \item \add{I/O accessing significantly degrades the performance of FHE applications}.
    \item \add{The use of multiple servers for distributed computing can not solve the problem and may even hurt application performance.}
    \item \add{The impact of I/O overhead can differ not only across various applications but also within the same application when different FHE parameters are used}.
\end{itemize}

We end this paper with a short summary and reflect on the possible future work if we intend to deploy FHE in the wild. We hope this paper can motivate interesting parties to further study storage in FHE, and thus commit to release the traces and software to the public.

\begin{figure}[t]
\centering
    \includegraphics[width=0.85\linewidth]{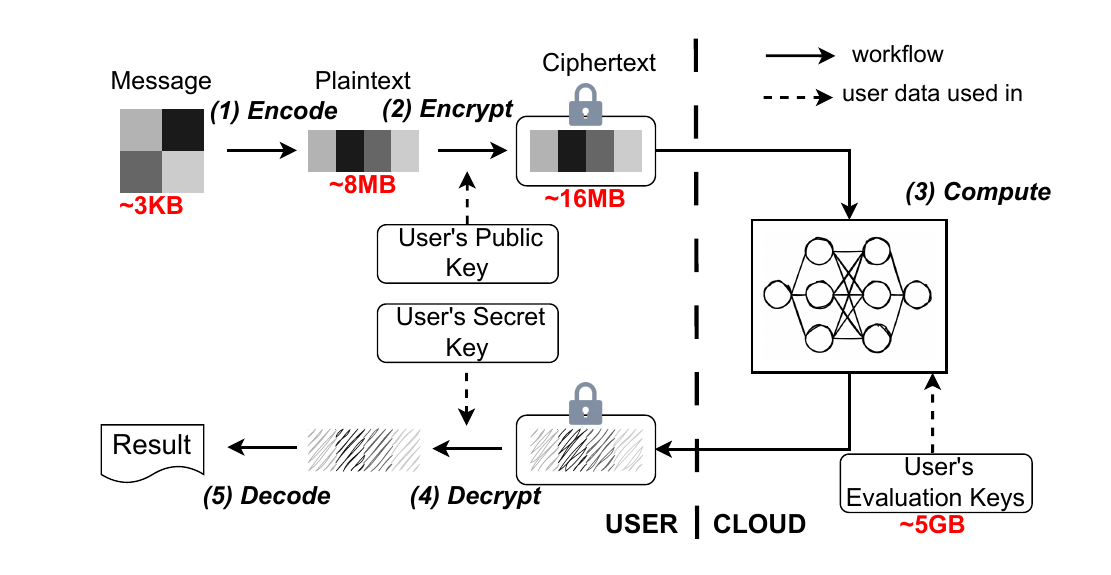} 
\caption{FHE workflow.\label{fig:fhe-dataflow}}
\vspace{-20pt}
\end{figure}

\section{FHE Workflow}\label{workflow}\label{sec:background}

Fully Homomorphic Encryption (FHE) is an emerging cryptographic technique that allows users to send encrypted data to a third party, like the public cloud, to perform operations. Afterwards, the users can locally decrypt the returned results as if the operations were performed on plain text. This technique helps users achieve efficient computation offloading and high data confidentiality at the same time. Currently, there are several FHE schemes that support different types of data, including BGV~\cite{bgv}, BFV~\cite{bfv1,bfv2}, TFHE~\cite{tfhe}, FHEW~\cite{fhew} and CKKS~\cite{ckks}. In this paper, we use the CKKS scheme because it supports floating-point numbers and has a wider application range~\cite{ckks,ckks-bootstrapping}.  In FHE, the end-to-end workflow consists of 5 steps: \textit{Encode}, \textit{Encrypt}, \textit{Compute}, \textit{Decrypt} and \textit{Decode}. Fig.~\ref{fig:fhe-dataflow} shows the workflow, and we use classifying an image with Resnet-20 (i.e., inference) to showcase the procedures of FHE with an emphasis on I/O. In this case, the user uploads an image to the server and the server return the classification result after computation.

\textbf{Encode.} First, FHE encodes the input (the image, a length-$n$ vector of pixels) as a polynomial with $N$ coefficients, satisfying $N/2 \ge n$. The goal is to \add{associate computations on polynomial rings with plain text computations}.
Then, the multiplication and addition of polynomials correspond to the element-wise multiplication and element-wise addition of the message vectors, respectively. Each coefficient of the encoded polynomial is an integer within the field modulo Q, where the modulus Q is a large integer typically exceeding 1000 bits in size~\cite{ckks-bootstrapping,sharp@isca23}. 

To reduce the high complexity of arithmetic computations between large integers, Chinese Remainder Theorem (CRT) is used to decompose a large integer in the modulo $Q$ domain into several small integers in the modulo $q_i$ domains, where the product of all $q_i$ is equal to $Q$, and each $q_i$ is a prime number that can be fit into the machine's word size. Thus, a large-coefficient polynomial can be represented as a series of small-coefficient polynomials, and each small-coefficient polynomial is called a $limb$.

In our case, the polynomial has $2^{16}$ coefficients. After encoding, an image of size 3KiB (dimensions 32$\times$32$\times$3) is transformed into a polynomial exceeding 8MiB in size.

\textbf{Encrypt.} Then, the user uses the public key to encrypt the plaintext version polynomial into a ciphertext, which consists of two polynomials and each has the same size as the plaintext polynomial. 
During the encryption process, the coefficients of the plaintext polynomial are combined with a random-generated error polynomial that guarantees the RLWE security~\cite{rlwe}.
In our case, after encryption, the 8MiB plaintext is converted into a 16MiB ciphertext.

\textbf{Compute.} In the \textit{Compute} step, the computation is composed of a series of primitive FHE operations over ciphertexts, including \textit{PAdd}, \textit{HAdd}, \textit{PMult}, \textit{HMult} and \textit{HRot}.
Table~\ref{table-he-ops} shows the primitive operations supported by the CKKS FHE scheme. 
Complex computations including deep neural networks~\cite{origin-resnet,mobilenet} and large language models~\cite{transformer,llama}, can be constructed by combining these primitive operations.  

The \textit{PAdd} operation performs the addition of a ciphertext and a plaintext, while the \textit{HAdd} operation performs the addition of two ciphertexts. These two operations are simple polynomial additions with a complexity of O($N$). 
The \textit{PMult} operation performs ciphertext-plaintext multiplication and the \textit{HMult} operation performs ciphertext-ciphertext multiplication. These two operations involve polynomial multiplication which have a complexity of O($N^2$). In implementation, the Number Theoretic Transform (NTT) technique is commonly used to accelerate polynomial multiplication. It transforms a polynomial from its coefficient representation to its point-value representation in O($N \log N$) time complexity. The multiplication of polynomials in point-value representation then has a complexity of O($N$). 
The \textit{HRot} operation can circularly left-shift elements of a ciphertext's underlying plain text vector. This operation is usually used to implement the accumulative operation of ciphertexts, e.g., matrix multiplication~\cite{ckks-bootstrapping,better-bootstrap}. Note that during the computation of \textit{HMult} and \textit{HRot} operations, auxiliary data, known as the evaluation key~\cite{ckks}, needs to be accessed. 
An evaluation key is composed of multiple ciphertexts that are independent of the input. 
All \textit{HMult} operations utilize the same evaluation key
while \textit{HRot} operations with different shift parameters correspond to different evaluation keys.  
In Resnet-20, there are more than a hundred different evaluation keys, with a total data size exceeding 5GiB.

\textbf{Decrypt and Decode.} The \textit{Decrypt} and \textit{Decode} are the inverse steps of encryption and encoding, respectively. During decryption, the resulting ciphertext is transformed into a plaintext polynomial, whereas in decoding, the plaintext polynomial is converted into the final result, which is a vector containing floating-point numbers.

\begin{table}[htbp]
\caption{\label{table-he-ops}CKKS primitive HE operations
}
\centering
\begin{tabularx}{\linewidth}{lX}
\hline
Operation & Description\\[0pt]
\hline
PAdd($C$, $P$) & Addition of a ciphertext $C$ and a plaintext $P$\\[0pt]
HAdd($C_1$, $C_2$) & Addition of two ciphertexts $C_1$ and $C_2$ \\[0pt]
PMult($C$, $P$) & Multiplication of a ciphertext and a plaintext\\[0pt]
HMult($C_1$, $C_2$, $\rm evk_{mult}$) & Multiplication of two ciphertexts $C_1$ and $C_2$. \\[0pt]
HRot($C$, $r$, $\rm evk_{rot}^r$) &  Circularly  left-shifts the elements of the ciphertext $C$'s underlying message vector by $r$ positions. \\[0pt]
\hline
\end{tabularx}
\end{table}

\section{Understanding IO Bottleneck in FHE}

\subsection{Storage I/O Overlooked\label{overlooked}}
In prior work on FHE, most works are focused on accelerating computation since \add{performing computations on ciphertext} can incur a heavy overhead (i.e., around $10^6\times$ more computation). While these GPU-/ASIC-based accelerators~\cite{tensorfhe@hpca23,gme@micro23,fab@hpca23,f1@micro21,mad@micro23,ark@micro22,bts@isca22,craterlake@isca22,sharp@isca23,fpga-rplp@hpca19,trinity@micro24,heap@isca24} significantly speed up the FHE computation (e.g., 2800$\times$ faster with TensorFHE~\cite{tensorfhe@hpca23} and $10^5\times$ faster with Sharp~\cite{sharp@isca23}), there is one aspect---the storage I/O---left understudied. 

Existing work operates under two assumptions: (1) all data (including ciphertexts and evaluation keys) is stored in high-bandwidth memory (HBM), and (2) the overhead associated with fetching data from HBM to on-chip cache can be eliminated through static optimal prefetching policy~\cite{fab@hpca23,craterlake@isca22}, data reuse algorithm optimization~\cite{ark@micro22,sharp@isca23,mad@micro23}, and a large on-chip cache \add{(200 to 500 MiB)}~\cite{sharp@isca23,craterlake@isca22,bts@isca22,ark@micro22,heax} which is able to accommodate the computing working set and the prefetched data.

However, these assumptions are not unlikely to be held in practice, especially in the context of cloud computing. First, existing  HBM has a limited capacity (around \add{tens of} \,GiB~\cite{a100-specs}). Recall that the FHE workflow can significantly enlarge the input (e.g., from a KiB-level image to MiB-level ciphertext, and GiB-level evaluation keys for running a Resnet-20 inference). \add{Note that different users would have their own ciphertexts and evaluation keys. Hence, it is unlikely to store all users' data inside HBM. Additionally, the application's input and temporary data can also occupy a significant amount of HBM space. For example, for a large language model with 13B parameter~\cite{opt-llm}, the model weight is 26GiB~\cite{vllm@sosp23}, and the KV cache~\cite{kv-cache} generated during processing a single request usually is 1.6GiB~\cite{vllm@sosp23}. For performance reasons, both the model weight and the KV cache need to be kept in HBM. Therefore, the available HBM space for FHE data is limited.}

Second, under ideal circumstances (i.e., with sufficient HBM memory, single application, and single user),  the static optimal data prefetching and scheduling policy~\cite{fab@hpca23,craterlake@isca22} can be realized to hide the data transfer overhead from HBM to the on-chip cache. Nevertheless, given the fact that the capacity of HBM is limited, data still has to be retrieved from external storage\add{, where the I/O overhead is difficult to hide}. An application may provide service for many users, and even though there may be multiple applications in a server competing with each other for hardware resources, it remains unclear to what extent the static policy can be effective. 

Third, although algorithm optimizations can reduce the working set of FHE applications and improve data reuse, these algorithm optimizations are only applicable to specific computational patterns. That is to say, they are only effective for some applications or certain computations within an application~\cite{ark@micro22}. For example, the algorithm optimization of ARK~\cite{ark@micro22} can effectively reduce the working set of the Resnet-20 application where the rotation operations have the same stride, but not for others (e.g., \add{HELR~\cite{helr} and Sorting~\cite{fhe-sorting} do not exhibit the similar pattern}). Note that even with these optimizations, an on-chip cache of several hundred megabytes is necessary to store the working set~\cite{ark@micro22,sharp@isca23}.

\subsection{Experiment Setup}
While the above have hinted the storage I/O can be impactful towards FHE overall performance, it remains unclear quantitatively. Hence, we conduct a series of experiments to demonstrate the variations in application performance under different setups including different types storage media and inter-connect networks.

\textbf{Baselines.} In this paper, we choose Sharp (the state-of-the-art ASIC accelerator)~\cite{sharp@isca23} and TensorFHE (the state-of-the-art GPU-based acceleration scheme)~\cite{tensorfhe@hpca23} as the two FHE accelerators used in our experiments. For Sharp, we use the same simulator described in its paper~\cite{sharp@isca23}. The baseline is the ideal performance reported in Sharp with the assumption that HBM is large enough and all the optimizations such as prefetching, operator scheduling, and data reuse are enabled. For TensorFHE, we use the public codes at~\cite{tensorfhe-docker} and run it with an NVIDIA A100 40 GB GPU and take the optimal performance as the baseline where all data are located in the GPU memory.

\textbf{Storage tiers.} We conduct simulation experiments on several storage tiers with different bandwidths, including HBM (1 TiB/s), 8-channel DDR5-5600 memory (358.4 GiB/s), PCIE5 $\times$16 Disk (64 GiB/s), and RDMA Disk (12.5 GiB/s). Since in FHE, the amount of data accessed at a time is usually dozens or hundreds of MiB, we only use throughput as the metric for evaluation. For test candidates, we intentionally choose to do cold start and bypass device's cache (i.e., no data in on-chip cache) to reflect the performance under multi-user setups like cloud.

\textbf{Applications.} Follow~\cite{ark@micro22,craterlake@isca22,sharp@isca23,tensorfhe@hpca23}, We conducted experiments on two of the most commonly used FHE applications, HELR~\cite{helr} and ResNet-20~\cite{fhe-resent}. HELR is a machine learning application that trains a binary classification model for the MNIST dataset~\cite{mnist} using logistic regression. The model is trained with a mini-batch containing 1024 images in each iteration. Similar to Sharp~\cite{sharp@isca23}, we train the model for 32 iterations, where an iteration is a gradient update step with a single batch, and report the average execution time per iteration. For ResNet-20, We performed CNN inference using the FHE CKKS implementation of the ResNet-20 model~\cite{origin-resnet} with an input image from the CIFAR-10 dataset~\cite{cifar10}.

\textbf{Distributed computing.} To simulate the distributed execution of FHE applications at scale, we use the cluster simulator SimGrid~\cite{simgrid}. SimGrid supports specifying the number of servers, the computing capabilities of each server, and the network topology among the servers. For Sharp, we integrate the SimGrid with the Sharp simulator where each server contains one Sharp accelerator. For TensorFHE, we do offline profiling for all the GPU kernels involved in distributed computing and import the profiling results into SimGrid to simulate the computational execution time when using multiple GPUs in distributed computing. 

\begin{figure}[htbp]
\centering
    \includegraphics[width=0.99\linewidth]{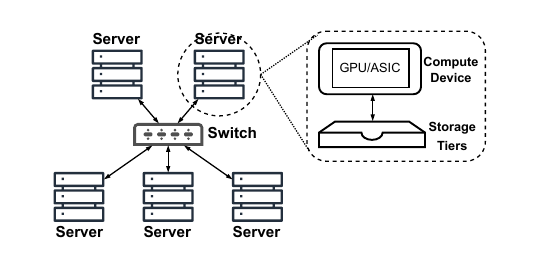} 
\caption{Experiment setup.\label{fig:exp-setup}}
\end{figure}

For the network topology among the servers, we adopt the star topology~\cite{star-topo} which has high performance and scalability. In the star topology, all servers are connected to the same central switch. We conducted experiments on three types of network links with different bandwidths to observe the impact of communication bandwidth on application performance, including Ethernet (400 Gb/s) and FastFabric (300 GiB/s). FastFabric refers to especially customized high-performance device-to-device interconnect links, such as NVIDIA NVLink~\cite{nvlink} and AMD Infinity Fabric~\cite{amd-if}. Similar to storage, we focus on evaluating bandwidth for FHE applications. Fig.~\ref{fig:exp-setup} illustrates the experimental setup comprising multiple servers interconnected via a switch. Each server houses a computing device (e.g., a GPU or an ASIC) that retrieve data from various storage tires.

In the experiments, we adopt the \textit{residue-polynomial-level parallelism} (rPLP) model~\cite{fpga-rplp@hpca19,f1@micro21} to partition the computational tasks among servers. In the rPLP model, a polynomial with large coefficients is represented by a series of small-coefficient residue polynomials~\cite{ckks}, and each server computes independent residue polynomials, which enables most operations to be computed locally on each server and avoids frequent communications~\cite{fpga-rplp@hpca19,f1@micro21}. 

\begin{figure}[t]
\centering

    \includegraphics[width=0.85\linewidth]{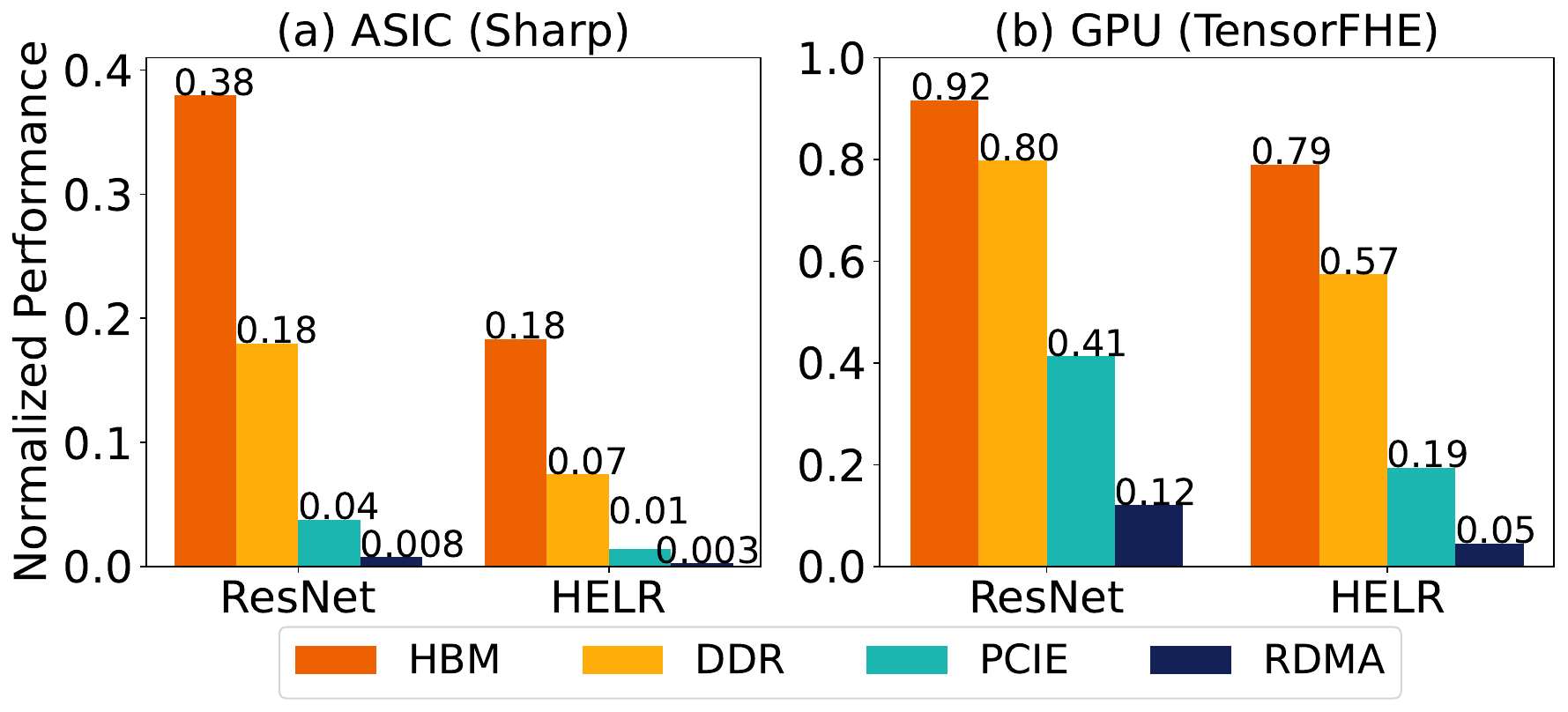} 
\caption{Performance under different storage: (a) ASIC, (b) GPU. The ASIC's baseline performance of ResNet-20 and HELR is 99ms and 2.5ms, and GPU's baseline performance is 4.9s and 220ms.  The I/O overhead significantly decrease the performance of ASIC and GPU.\label{fig:all-storage-perf}}
\end{figure}

\section{Observation}

\subsection{Locality}

\begin{figure}[htbp]
\centering

    \includegraphics[width=0.85\linewidth]{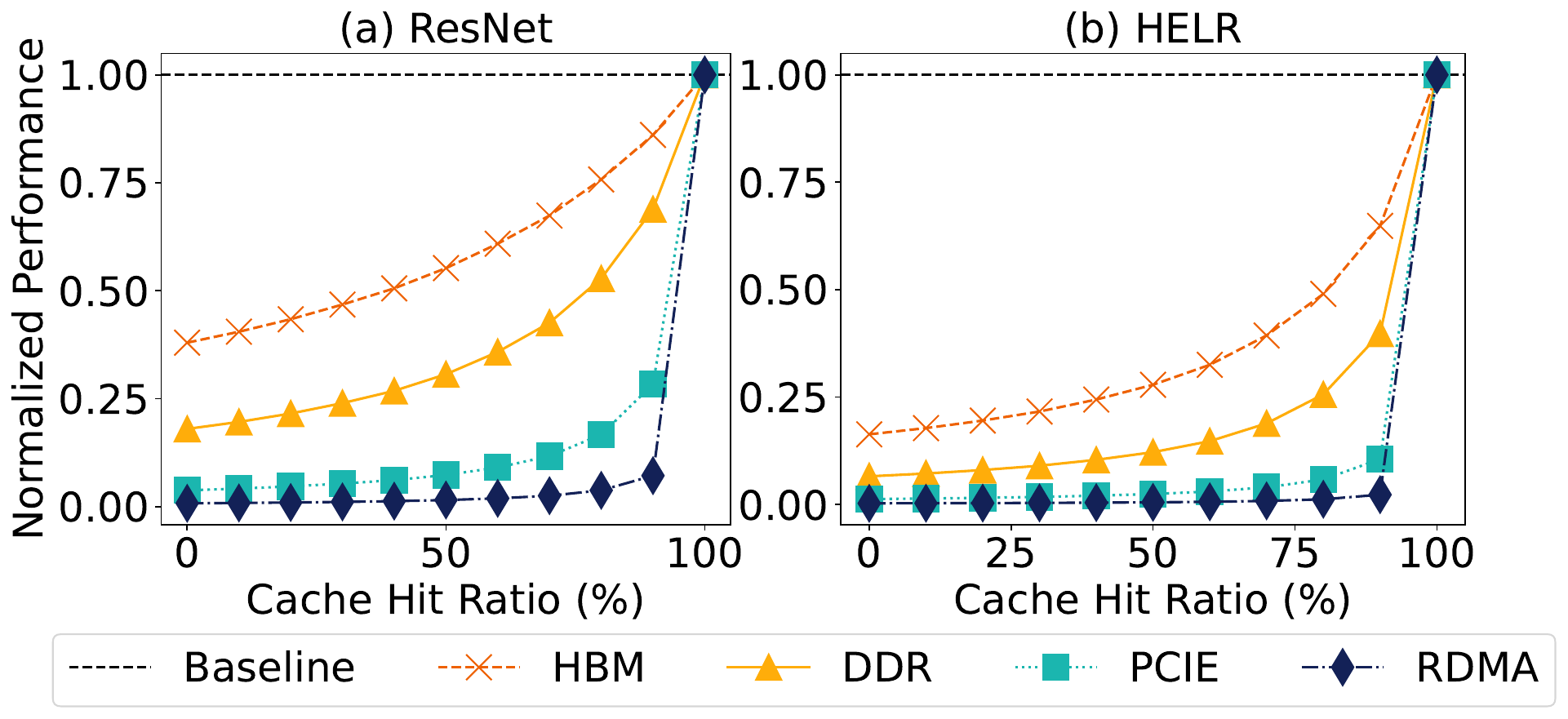} 
\caption{Performance under varying cache hit ratio: (a) Resnet-20 (b) HELR. Different 
storages and different applications need different cache hit ratios to achieve a fixed target performance.\label{fig:cache-hit-ratio}}
\end{figure}

\textbf{Observation.} As shown in Fig.~\ref{fig:all-storage-perf}, compared to the Sharp baseline (which implements a statically optimal replacement strategy to focus on accessing on-chip memory), application performance significantly degrades if each evaluation key must be retrieved from off-chip storage. Specifically, the performance will have an average 4.0$\times$ slowdown even if the off-chip storage is HBM, and it can reach up to 244$\times$ if the off-chip storage is an RDMA disk. This observation also applies to the GPU. Although the GPU performance only slightly decreases when accessing evaluation keys from HBM or DDR5 (1.2$\times$ slowdown and 1.5$\times$ slowdown respectively), when it needs to access data from PCIE disk or RDMA disk, the performance slowdown will be 3.8$\times$ and 15.2$\times$.

\textbf{Root cause.}  The significant performance decline of Sharp can be attributed to its extremely high I/O pressure, averaging 3381 bytes per cycle. In contrast, TensorFHE experiences relatively lower I/O pressure thanks to lower processing capability with 101 bytes per cycle on average. Considering the substantial on-chip cache available in ASIC accelerators, we further analyze the impact of on-chip cache hit ratios on performance. For Sharp, we manually set the hit ratio of the on-chip cache from 0 to 100\%, and then evaluate the performance changes. In Fig.~\ref{fig:cache-hit-ratio}, for a specific performance, the lower the storage bandwidth is, the higher the hit ratio that needs to be achieved. For example, to reach 80\% of the performance of Resnet-20 and HELR baselines, for HBM, DDR, PCIE, and RDMA, the average cache hit ratios that need to be achieved are 90.2\%, 96.2\%, 99.3\%, and 99.9\% respectively.

\textbf{Takeaway.} I/O accessing significantly degrades the performance of FHE accelerators. 
To deploy FHE in the field, especially as a cloud service, handling storage would be, if not already is, the bottleneck of the overall performance.

\subsection{Distributed Computing}
\begin{figure}[htbp]
\vspace{-25pt}
\centering

    \includegraphics[width=0.75\linewidth]{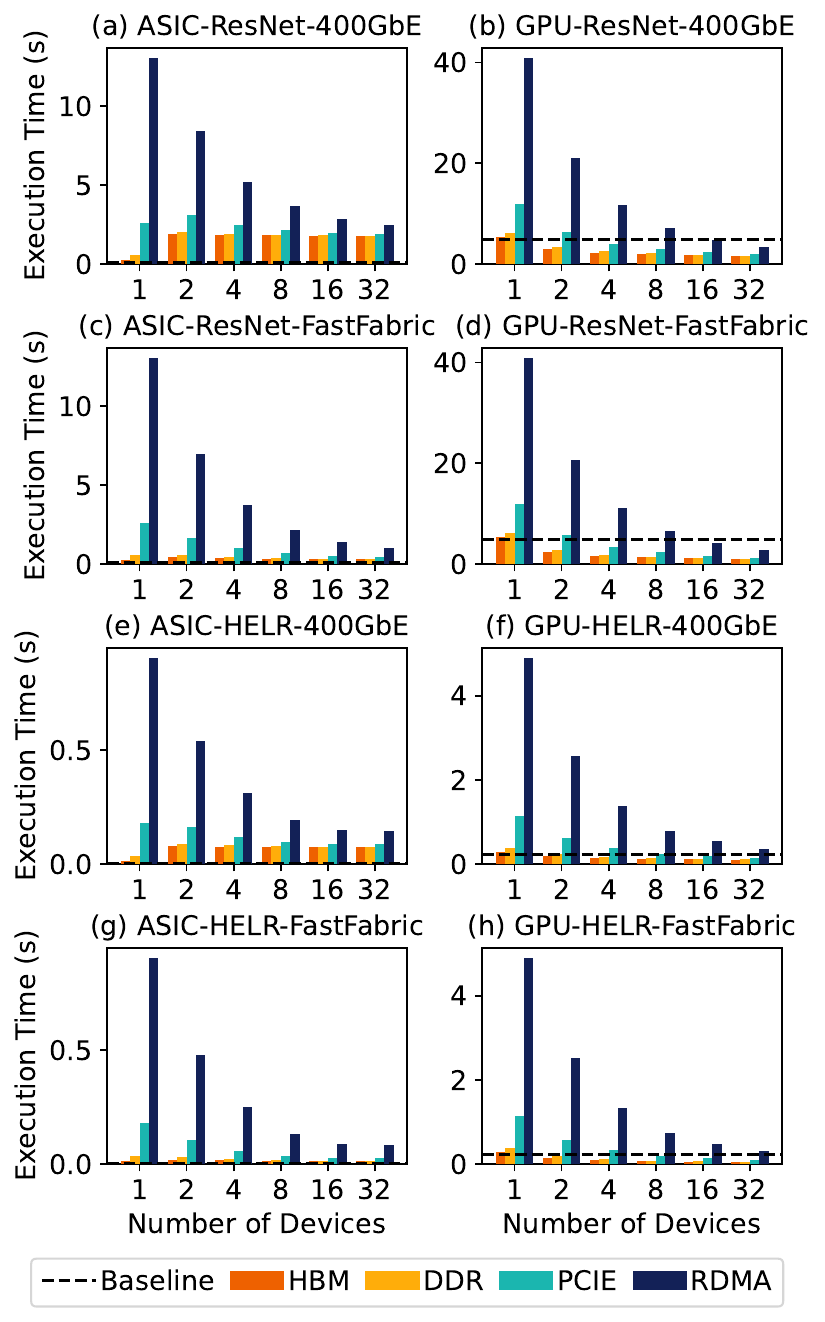} 
\caption{Performance variation with different accelerators, storages, networks, and applications.\label{fig:all-distributed-perf}}
\vspace{-20pt}
\end{figure}

\textbf{Observation.} Fig.~\ref{fig:all-distributed-perf} shows the variation of applications' execution time with the number of distributed hosts when using Ethernet and FastFabric as the interconnect networks between hosts. For TensorFHE, distributed computing can effectively improve performance. Compared to single-host computation, using 32 hosts results in an average performance speedup of 6.6$\times$ (Ethernet) and 9.7$\times$ (FastFabric).
However, for Sharp, distributed computing can lead to a decrease in performance under certain conditions. When using Ethernet, for PCIE and RDMA, compared to single-host computing, using 32 hosts results in average speedups of  1.72$\times$ and 5.78$\times$, respectively. But for HBM and DDR, using 32 hosts results in slowdowns of 6.08$\times$ and 2.74$\times$. The slowdowns are mainly caused by the communication overhead. When using FastFabric, the average speedups are 0.94$\times$, 1.99$\times$, 6.42$\times$ and 11.96$\times$, for HBM, DDR, PCIE and RDMA, respectively.

\begin{figure}[htbp]
\centering
 \includegraphics[width=0.85\linewidth]{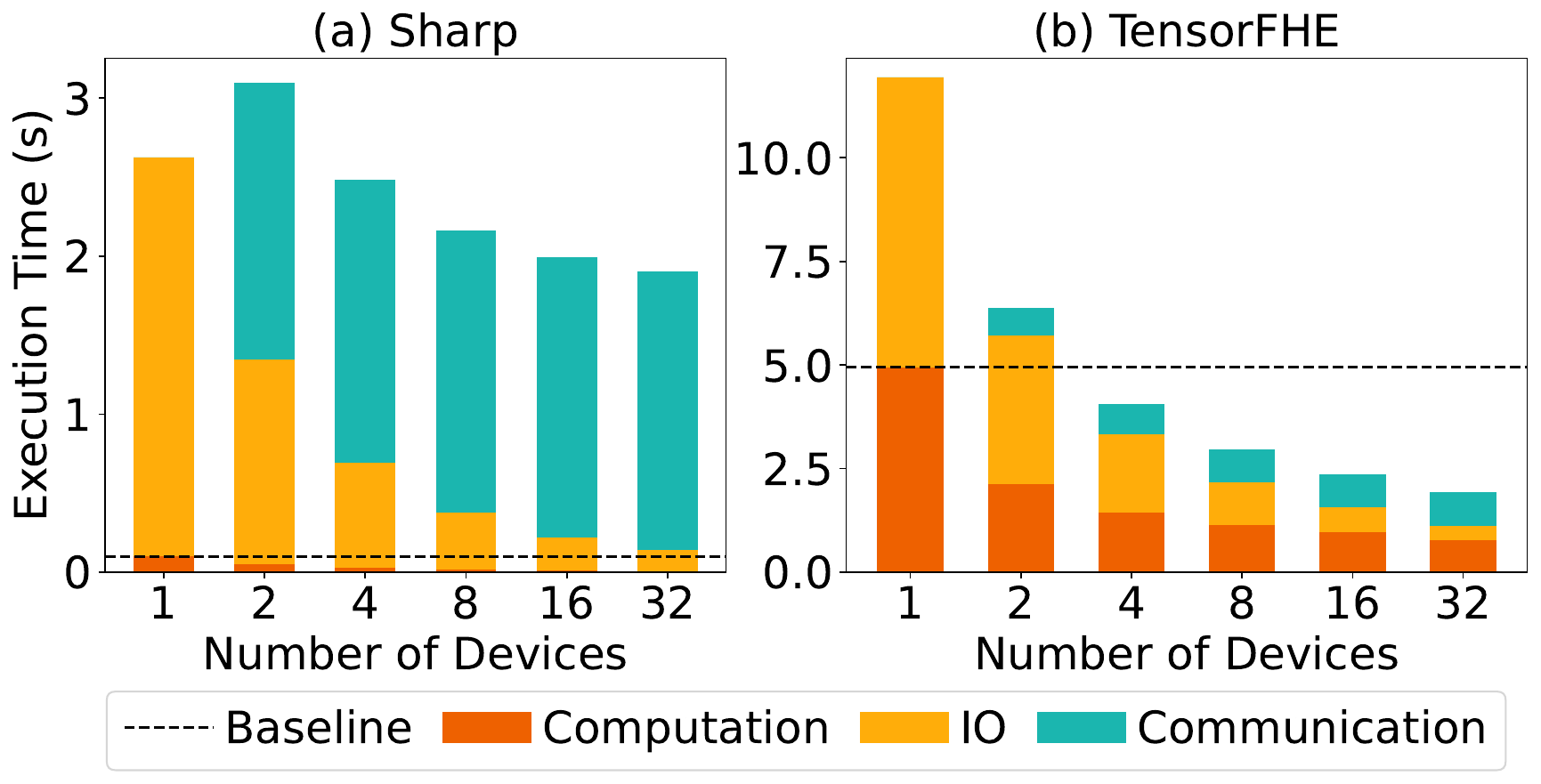} 
\caption{Execution time breakdown: (a) Sharp (b) TensorFHE. 
The communication overhead became the dominant overhead in Sharp, and computation overhead and communication overhead are the main bottlenecks in TensorFHE.\label{fig:exec-time-breakdown}}
\end{figure}

\textbf{Root cause.}  For Sharp and TensorFHE, distributed computing can both reduce the computational overhead and IO overhead, but the communication overhead instead becomes the new bottleneck. We show the execution time breakdown of Resnet-20 when using PCIE as the storage and Ethernet as the network in Fig.~\ref{fig:exec-time-breakdown}, a common setup in the field. The conclusion is also applicable to other types of storage, network configurations, and HELR application.  For Sharp, as the number of hosts increases, the computational overhead and I/O overhead decrease proportionally, reducing from 3.8\% and 96.2\% with 1 host to 0.3\% and 7.2\% with 32 hosts, respectively. The communication overhead becomes the bottleneck, reaching a maximum proportion of 92.5\% with 32 hosts.
For TensorFHE, as the number of hosts increases, the storage overhead decreases proportionally, but the computing overhead decreases slowly.
This is because the architecture of GPUs is well-suited for batch processing~\cite{gpgpu}, a reduction in the amount of computation does not lead to a proportional decrease in computation time.  For TensorFHE, when using 32 hosts, the proportions of computation overhead, I/O overhead, and communication overhead are 40.1\%, 18.1\%, and 41.8\%, respectively. 

\textbf{Takeaway.} For ASIC, as the number of hosts increases, the computational overhead and I/O overhead decrease proportionally, and the communication overhead becomes the dominant factor in limiting the performance, while for GPU, both the communication and computation overhead are still the main bottlenecks.

\subsection{Applications}

\textbf{Observation.}   For different applications, the impact of I/O overhead on performance is different. As shown in Fig.~\ref{fig:all-storage-perf}, when considering I/O overhead, the HELR application experiences a higher slowdown compared to the ResNet-20 application. For Sharp, the performance slowdown for ResNet-20 on HBM, DDR, PCIE, and RDMA is 2.63$\times$, 5.56$\times$, 26.5$\times$, and 131.7$\times$, respectively. In contrast, the performance slowdown for HELR increases to 5.5$\times$, 13.4$\times$, 70.6$\times$, and 357.2$\times$, respectively. 

For the same application, using different FHE parameters will affect the I/O overhead. For instance, Sharp uses 1555-bit integers as polynomial coefficients, while TensorFHE uses 840-bit integers and 1092-bit integers for ResNet-20 and HELR applications, respectively. This means that for Resnet-20 and HELR applications, the size of TensorFHE's polynomials is only 54\% and 70\% of the size of Sharp's polynomials, respectively. But on the other hand, \add{to get the best performance, TensorFHE has selected a different FHE parameter set suitable for GPU~\cite{tensorfhe@hpca23}, which makes the average size of each evaluation key in TensorFHE 5.5$\times$ the size of that in Sharp.}  Additionally, different FHE parameters correspond to different combinations of fundamental FHE operators (such as the number of rotations), which in turn affect the I/O demand. In experiments, we observed that for Resnet-20 and HELR, the total I/O data volume of TensorFHE is 2.8$\times$ and 4.5$\times$ of Sharp's, respectively.

\textbf{Root cause.} Different applications use different combinations of fundamental FHE operators, influencing the data access patterns. For example, the HELR application involves a large number of rotation operations to implement inner products of vectors~\cite{helr}, leading to significant I/O demand for evaluation keys. In contrast, the Resnet-20 application consists of a lot of addition operations and ciphertext-plaintext multiplication operations, which only require access to the corresponding ciphertext and plaintext polynomials, not the evaluation keys, thus resulting in a lower I/O demand. Specifically, the Sharp Resnet-20 baseline needs to read 1633 bytes of evaluation keys on average per cycle, while the HELR baseline needs to read 5130 bytes on average per cycle, which is 3.1$\times$ of Resnet-20. For a same application, different FHE parameters correspond to different security level, data precision, and computation procedure, which not only affect the size of ciphertexts and evaluation keys, but also affect the number of FHE operations.

\textbf{Takeaway.} The impact of I/O overhead can differ not only across various applications but also within the same application when different FHE parameters are used.

\section{Future Directions}
Through our experiments, we reveal that the performance of FHE under existing hardware and distributed computing paradigm are still far from deployment in practice. 
Here, we list a few possible directions in the hope that they shall motivate further research in this domain.

\textbf{Locality-first scheduling.}
Since distributing the FHE among nodes is not always helpful due to hardware limited throughput/bandwidth, it would be ideal to have dedicated servers for users to reduce the I/O accessing of FHE data. However, this undoubtedly leads to low resource efficiency. A possible solution is to investigate user access patterns and pipeline the access to hide the context switch overhead.

${\textbf{Near-data processing.}}$ Previous studies have demonstrated that near-data processing can effectively reduce I/O overhead ~\cite{pact-ndp,database-ndp,micro-ndp}. By integrating FHE computation components into storage devices, FHE computations that require extensive I/O can be offloaded to the storage devices, thereby minimizing the I/O between the storage devices and the host. This represents the most promising direction, as the evaluation keys in FHE are only accessed during specific FHE operations (e.g., $HRot$ and $HMult$). Therefore, it is feasible to design dedicated computing units specifically for these operations and integrate them into storage devices. By adopting this approach, the I/O overhead associated with evaluation keys can be significantly reduced.

\textbf{I/O-friendly application implementation.} Combining the characteristics of FHE I/O to optimize application implementations can also effectively reduce the I/O overhead. For example, FHE addition does not require reading the evaluation key. Therefore, we can exploit this feature to 
refactor the program. Such changes can be counter-intuitive (e.g., increasing computation overhead) at first sight but would lead to better overall performance, especially given the fast growing processing ability of FHE accelerators.

%
%

\bibliographystyle{splncs04}
\bibliography{paper}





\end{document}